\documentclass[usenatbib]{mn2e}
\usepackage{graphicx}

\title[Photons Delay Fluctuation Time]{Fluctuations of Photon Arrival Times in Free Atmosphere}

\author[S. Cavazzani et al.]{S. Cavazzani$^{1}$\thanks{E-mail:stefano.cavazzani@unipd.it}, S. Ortolani$^{1}$, C.Barbieri$^{1}$\\
$^{1}$Department of Astronomy, University of Padova, Vicolo
dell'Osservatorio 3, I-35122, Padova, Italy\\}

\begin{document}

\date{Accepted 2011 September 23. Received 2011 September 12; in original form 0000 Month 00}

\pagerange{\pageref{firstpage}--\pageref{lastpage}} \pubyear{2011}

\maketitle

\label{firstpage}

\begin{abstract}

In this paper we calculate the delay of the arrival times of visible photons on the focal plane of a telescope and its fluctuations as function of local atmospheric conditions (temperature, pressure, chemical composition, seeing values) and telescope diameter. The aim is to provide a model for delay and its fluctuations accurate to the picosecond level, as required by several very high time resolution astrophysical applications, such as comparison of radio and optical data on Giant Radio Bursts from optical pulsars, and Hanbury Brown Twiss Intensity Interferometry with Cerenkov light detectors. The results here presented have been calculated for the ESO telescopes in Chile (NTT, VLT, E-ELT), but the model can be easily applied to other sites and telescope diameters. Finally, we describe a theoretical mathematical model for calculating the Fried radius through the study of delay time fluctuations.

\end{abstract}


\section{Introduction}

Several high time resolution instruments are able to measure the arrival time of visible photons with an internal precision in the range 10 - 100 picoseconds, and refer those arrivals to the commonly used UTC scale with an accuracy of the order of 500 picoseconds (see for instance, Barbieri et al. \cite{barbieri} and Naletto \& Barbieri, \cite{naletto}).\\
The step of referring the arrival times to UTC is usually done with the intermediary of time signals broadcasted over radio frequencies (e.g. GPS, GLONASS, Galileo GNSS), signals which are very accurately corrected for atmospheric propagation effects like ionospheric scintillation and wet troposphere refraction. In the usual astronomical applications, the arrival times of optical photons are not corrected to the same accuracy.\\
The possibility to perform such correction is actually shown by accurate laser ranging to geodetic satellites (recall that 1 nanosecond in vacuum corresponds to 30 cm).
For instance, Kral et al. (\cite{ivan}) quote a precision of few picoseconds by taking into account the atmospheric seeing. Moreover a recent discussion has been performed by Dudy D.Wijaya and Fritz K. Brunner (\cite{wijaya}) on the atmospheric range correction for two-frequency Satellite Laser Ranging (SLR) measurements.\\
Motivated by our own very precise time measurements on celestial sources with Aqueye and Iqueye (\cite{barbieri} and Naletto \& Barbieri, \cite{naletto}), we have undertaken the calculation of the delay and delay dispersion of visible photon arrival times in the usual conditions prevailing in astronomical observatories. 
In the first step of our procedure, the Marini-Murray model (Marini \& Murray, \cite{marini}) is used to calculate the correction $\Delta R$ to the optical path of photons in air.\\
Through this calculation, the atmospheric refractive index $n$ and a fixed delay time independent of the photometric night quality is derived. Then the photon paths are correlated with the astronomical seeing.\\
Through this relation we derive a statistical set of delay times as function of the Fried radius $r_{0}$ and telescope diameter. Finally, the difference of these delay times with the fixed delay gives the fluctuations.\\
Reversing this model we also developed a theoretical mathematical model for the $r_{0}$ calculating through the observation of these fluctuations.\\
The results are expounded in the present paper, which is organized as follows:

\begin{itemize}
	\item Section 2 describes the modified Marini-Murray model
	\item In Section 3 the refraction index and delay time is calculated 
	\item In Section 4 the Fried radius is introduced and the seeing effects on the image are recalled
	\item In Section 5 the geometric and physical delay time fluctuations is introduced 
	\item In Section 6 we describe the model for the total delay time fluctuations
	\item In Section 7 we calculate $r_{0}$ through the study of delay time fluctuations
\end{itemize}

\section{MARINI-MURRAY Model for the Refraction Correction}

\begin{figure}
  \centering
  \includegraphics[width=8.5cm]{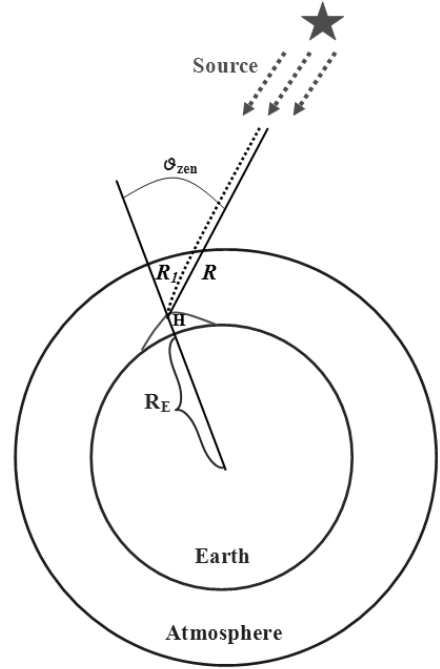}
  \caption{Optical path length (OPL) $R$ and its variation $\Delta R$ due to atmospheric refraction. $R$ is the OPL in vacuum, $R_{1}$ is the OPL in atmosphere, $R_{E}$ is the Earth radius which the model calculates according to the latitude, $\theta_{zen}$ is the Zenith angle and $H$ is the site altitude.}
             \label{ref}
   \end{figure}

The Marini-Murray model is based on an expansion of hypergeometric functions (Marini \& Murray, \cite{marini}).
The model relies on hydrostatic equilibrium and the barometric equation. It considers a hydrostatic water vapor distribution and the water vapor behavior such as that of an ideal gas.
The refraction correction is given as a function of temperature, pressure, vapor partial pressure, gravity acceleration and universal gas constant.
The model also includes the values of  water vapor molar mass and air average molar mass.
Finally the refraction correction is calculated in relation to the Earth radius, the site altitude and latitude.
Through these parameters we can determine Optical Path Length (OPL) correction as described in detail below.
According to the original Marini-Murray model, the refraction correction $\Delta R = R_{1}-R$ (see Fig. \ref{ref}), is calculated as:

	\[\Delta R=\frac{f(\lambda)}{g(\phi,H)}\left[\frac{g_{1}+g_{2}+g_{3}}{sin(\theta_{\omega})+\frac{\frac{g_{2}}{g_{1}+g_{2}+g_{3}}}{sin(\theta_{\omega})+0.01}}\right]	
\]

where $g_{1}$, $g_{2}$, $g_{3}$ and $g(\phi,H)$ are defined below, $\theta_{\omega}$ is angle of elevation, $\phi$ is the latitude and $H$ is the altitude of the observatory.
The several terms are defined as:

\begin{itemize}

\item \[g_{1}=80.343\cdot 10^{-6}\left[\frac{R_{G}}{M_{d}\cdot\vec{g}}P+(1-\frac{M_{\omega}}{M_{d}})\frac{R_{G}}{4M_{d}\cdot\vec{g}}P_{\omega}\right]
\]

\item \[g_{2}=10^{-6}\left[\frac{80.343\cdot R_{G}}{R_{E}\cdot M^{2}_{d}\cdot\vec{g}^{2}}P\cdot T\cdot K(\phi,T,P)\right]+
\]
	\[+10^{-6}\left[10^{-12}\frac{80.343^{2}\cdot 2R_{G}\cdot P^{2}}{4M_{d}\cdot\vec{g}\cdot T(3-\frac{1}{K(\phi,T,P)})}\right]
\]

\item \[g_{3}=-10^{-6}\left[\frac{11.3\cdot R_{G}}{g(\phi,H)\cdot 4M_{d}\cdot\vec{g}}P_{\omega}\right]
\]

\end{itemize}

where:

\begin{itemize}

\item \[g(\phi, H)=1-0.0026\cdot cos(2\phi)-0.00031\cdot H
\]

\item \[K(\phi,T,P)=1.163-0.00968\cdot cos(2\phi)-
\]

	\[+0.00104\cdot T-0.00001435\cdot P
\]

\end{itemize}

In the original model the function $f(\lambda)$ was defined by with the following formula:

	\[f(\lambda)=0.965+\frac{0.0164}{\lambda^{2}}+\frac{0.000228}{\lambda^{4}}
\]

However, the refraction index formula used in the original Marini-Murray formalism is valid for a limited wavelength range ($0.40\mu m \div 0.60\mu m$). This is due to the model assumptions. For this reason, Ciddor (\cite{ciddor}) devised a more refined model valid for a wider wavelength range ($0.30\mu m \div 2.00\mu m$).
This refined model is described by the formula (Riepl et al. \cite{riepl}):

\begin{equation}
\label{rie}
\Delta R=f_{Gr}(\lambda)\left[\frac{g_{1}+g_{2}}{sin(\theta_{\omega})+\frac{\frac{g_{2}}{g_{1}+g_{2}}}{sin(\theta_{\omega})+0.01}}+\frac{g_{3}}{sin(\theta_{\omega})}\right]	
\end{equation}

where the dispersion formula, taken from Ciddor (\cite{ciddor}) normalized to the wavelength of $0.6943\mu m$, gives:

\begin{equation}
	f_{Gr}(\lambda)=\frac{k_{1}(k_{0}+(\frac{1}{\lambda})^{2})}{(k_{0}-(\frac{1}{\lambda})^{2})^{2}}+\frac{k_{3}(k_{2}+(\frac{1}{\lambda})^{2})}{(k_{2}-(\frac{1}{\lambda})^{2})^{2}}
\end{equation}

The values of the involved a-dimensional constants resulting from the normalized dispersion formula are:\\
$k_{0}$=238.0185, $k_{1}$=205.0638, $k_{2}$=57.362 and $k_{3}$=5.944936.\\
In the formulae, $T\left[K\right]$ is the temperature, $P\left[mb\right]$ is the total pressure, and $P_{\omega}\left[mb\right]$ is the water vapor partial pressure,  $\vec{g}=9.784\left[\frac{m}{s^{2}}\right]$ is the acceleration of gravity at the equator. The gas constant is taken as $R_{G}=8314.36\left[\frac{mJ}{K}\cdot mol\right]$.
The following molecular values are used:

\begin{itemize}

	\item $M_{\omega}=18.016\left[\frac{g}{mol}\right]\Rightarrow$ Water vapor
	\item $M_{d}=28.966\left[\frac{g}{mol}\right]\Rightarrow$ Air average.
	
\end{itemize}

Equation \ref{rie} fills the Marini-Murray simplifications with a hybrid approach in which the dry and wet refraction delays are treated
separately.

\section{Refractive Index and Delay Time Calculation}

Through the refraction correction path $\Delta R$ we can calculate, in a first approximation, the mean atmospheric refractive index $n_{1}$.
In fact, we note that the $n_{1}$ fluctuation is of the order of $10^{-9}$ for 1 ps (see Section \ref{pdtf}).
If we assume that $n_{1}$ has a constant distribution:

\begin{equation}
\label{index}
	n_{1}=\frac{OPL+\Delta R}{OPL}
\end{equation}

where the geometric optical path length (OPL) is given by the formula:

\begin{equation}
\label{opl}
	OPL=\frac{AA-SA}{cos(\theta_{zen})}
\end{equation}

where:

\begin{itemize}

	\item AA=Atmosphere indicative thickness (In this paper we have considered the $AA=10Km$)
	\item SA=Site altitude
	
\end{itemize}

The refractive angle $\theta_{1}$ is then calculated as:

	\[\theta_{1}=arcsin\left(\frac{1}{n_{1}}\right)
\]

With the value of the refractive index $n_{1}$ the delay time $t_{1}$ is calculated through the formula:

\begin{equation}
\label{delay7}
	t_{1}=\frac{n_{1}\cdot OPL}{c/n_{1}}-\frac{OPL}{c}
\end{equation}

where $c$ is the velocity of light in vacuum $\left(c=299792.458\left[\frac{Km}{s}\right]\right)$.
Equation \ref{delay7} allows us to calculate the propagation delay time in the atmosphere as a function of the refractive index $n$, including the geometric effects.\\
The first term in Equation \ref{delay7} is the ratio between the geometric OPL and the speed of light in the atmosphere. The main contribution of the delay time comes from the denominator of this ratio.

\subsection{Application of the model to ESO astronomical sites for the delay time}

In this Section we calculate the delay time for the three Chilean sites of ESO telescopes. Table \ref{st} shows the sites characteristics. We assume an average ground temperature of 288 K and an average ground relative humidity of 20\% for the three sites. Table \ref{delay} shows the obtained values from the simulation (through the Equation \ref{delay7}) at different Zenith angles. The delay time is calculated as a function of wavelength ($\lambda=0.632\mu m$) and has been calculated for a zenith angle ranging from $0^{\circ}$ to $60^{\circ}$. The data show the increase of the delay time with the $\theta_{zen}$ following the trend of the $cos$ function. We note that this delay does not depend on the telescope diameter and has variations in the order of $10^{-2}ns$ between Paranal and La Silla. These variations are due to the difference in altitude and geographical coordinates of the sites.

\section {Calculation of the Fried parameter and Seeing}

The atmospheric optical turbulence introduces variations on $t_{1}$ according to refraction fluctuations. 
We now recall one of the main parameters of the astronomical seeing, namely the Fried's radius $r_{0}$, which defines the average size of the turbulent cell. This parameter will be useful to introduce the concept of photons delay time fluctuations.
Fried has shown (Fried, \cite{fried}), within the limits of validity of the Kolmogoroff law, that $r_{0}$ is expressed by the formula:

\begin{equation}
	r_{0}=\left[0.423\cdot\frac{4\pi^{2}}{\lambda^{2}}\cdot\frac{1}{cos(\theta_{zen})}\int C^{2}_{n}\cdot dz\right]^{-\frac{3}{5}}
\end{equation}

where $C^{2}_{n}$ is the refractive index structure parameter:

	\[C^{2}_{n}=\left[80\cdot10^{-6}\frac{P}{T}\right]\cdot C^{2}_{T}
\]

and the temperature structure parameter $C^{2}_{T}(x)$ is defined through the formula:

\[C^{2}_{T}(x)=\frac{\left[T(x)-T(x+\Delta x)\right]^{2}}{\Delta x^{-\frac{2}{3}}} 
\]

this parameter is expressed in $\left(^{\circ}C\right)^{2}\cdot m^{-\frac{2}{3}}$ and expresses the temperature variations between two locations at a distance $\Delta x$.

\subsection{Seeing effects on the images}

The seeing produces scintillation, smearing and motion of the image.
Roddier (Roddier, \cite{roddier}) has obtained the following approximate expressions for the calculation of these three effects.

\begin{table}
 \centering
 \begin{minipage}{80mm}
  \caption{Geographic Characteristics of the Sites.}
   \label{st}
  \begin{tabular}{@{}lcccc@{}}
  \hline

  Site      &LAT.&      LONG. & Altitude & Telescope Diameter \\
            &    &            &  Km      &  m         \\
 \hline
 
 La Silla   &   $-29^{\circ}15'$  &  $-70^{\circ}43'$  &$2.347$     &  $3.58$    \\
 Paranal    &   $-24^{\circ}37'$  &  $-70^{\circ}24'$  &  $2.630$   &  $8.20$  \\
 Armazones   &   $-24^{\circ}35'$  &  $-70^{\circ}11'$  &  $3.064$   &  $42.00$   \\
 
 \hline

\end{tabular}
\end{minipage}
\end{table}

\subsubsection{Scintillation}

The image scintillation, as a function of $C^{2}_{n}$, in approximation is given by the following formula:

	\[\frac{\sigma^{2}_{I}}{I}\propto D^{-\frac{7}{3}}\cdot \frac{1}{(cos(\theta_{zen}))^{3}}\cdot \int C^{2}_{n}(z)\cdot z^{2}\cdot dz
\]

where $D$ is the diameter of the telescope. Scintillation however does not affect the dispersion of the arrival times, and therefore is not taken further into account.

\begin{table}
 \centering
 \begin{minipage}{80mm}
 \caption{Delay time vs Zenith angle for different sites ($\lambda=0.632\mu m$). 
 We note that this delay time does not depend on the telescope diameter and varies slightly with the location of the site.}

  \label{delay}
  \begin{tabular}{@{}llllcccccccc@{}}
  \hline
 Zenith Angle & $\left(^{\circ}\right)$ & 0 & 15 & 30 & 45 & 60 \\

 \hline
 La Silla   & (ns)  & 15.53 & 16.24 & 18.58 & 23.14 & 30.72 \\
 Paranal    & (ns) & 15.52 & 16.23 & 18.57 & 23.13 & 30.71  \\
 Armazones  & (ns)  & 15.52 & 16.23 & 18.57 & 23.13 & 30.71 \\
 \hline

\end{tabular}
\end{minipage}
\end{table}

\subsubsection{Image Smearing}

The light from the point source is spread over an area having a Full Width Half Maximum (FWHM) given by:

	\[FWHM=0.98\frac{\lambda}{r_{0}}
\]

The value is in $arcsec$. The amplitude of this effect is independent of the pupil diameter.

\subsubsection{Image Motion}

The motion of the image, as a function of $\lambda$, telescope diameter (D) and $r_{0}$, is given by:

\begin{equation}
	\sigma^{2}(x)=\sigma^{2}(y)=0.18\cdot \lambda^{2} \cdot D^{-\frac{1}{3}}\cdot r^{-\frac{5}{3}}_{0}
\end{equation}

This motion is in $arcsec$, and can be expressed in linear units ($meters$) by means of the formulae:

	\[\Delta x=n_{1}\cdot OPL\cdot sin \sqrt{\sigma^{2}(x)}
\]

	\[\Delta y=n_{1}\cdot OPL\cdot sin \sqrt{\sigma^{2}(y)}
\]

\begin{figure}
  \centering
  \includegraphics[width=8.5cm]{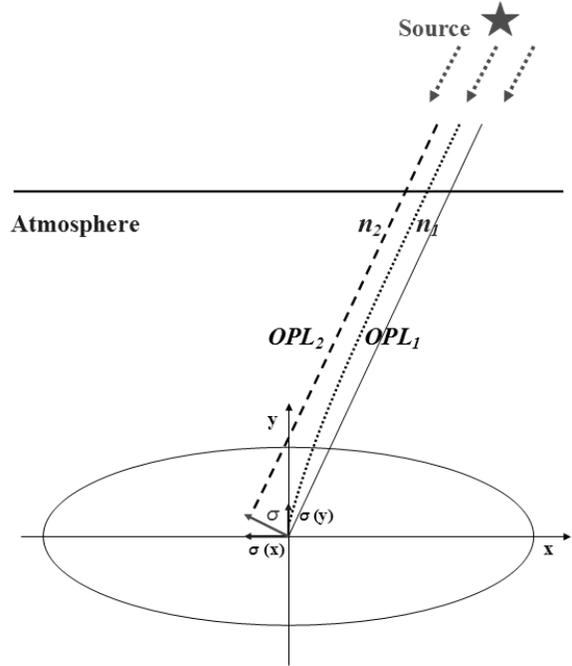}
  \caption{Schematic diagram of the OPL variation due to the image motion. Figure shows the optical path length in vacuum and the graphical representation (not to scale) of its variations. We define $OPL_{1}=n_{1}\cdot OPL$ and $OPL_{2}=n_{2}\cdot OPL$. Considering the motion of the image $\sigma<<r_{0}$ we can assume that relevant $x$ and $y$ refractive index gradients are unexpected.}
             \label{mot}
   \end{figure}

The latter effect is very important for the fluctuation of delay times.
One can calculate a new optical path as a function of this image motion, and then a new delay time (See Figure \ref{mot}). Figure \ref{mot} shows the optical path length in vacuum and the graphical representation (not to scale) of its variations.\\
We define $OPL_{1}=n_{1}\cdot OPL$ and $OPL_{2}=n_{2}\cdot OPL$, where $n_{1}$ is calculated using the Marini-Murray model (see Formula \ref{index}) and $OPL$ through the Formula \ref{opl}.
The difference between the previously calculated time and this new one gives the fluctuation.
The total motion is given by the formula:

\begin{equation}
	\Delta=\sqrt{\Delta^{2}x+\Delta^{2}y}
\end{equation}

This motion is $<<OPL$, then we can calculate the new OPL through the formula:

\begin{equation}
	OPL_{2}=\sqrt{OPL^{2}_{1}+\Delta^{2}}
\end{equation}

where:

	\[OPL_{1}=n_{1}\cdot OPL.
\]

\section{Geometric and Physical Delay Time Fluctuation}
\label{gf}

In this Section we calculate the delay time fluctuation. It is due to two contributions:
one correction is due to the change in photon path (geometric), and the other (physical) is due to the fact that
the photon is traveling in a medium where the refractive index is changed.
The delay time ($t_{2,G}(r_{0},D)$) due only to the geometric variation of the OPL is given by:

	\[t_{2,G}(r_{0},D)=\frac{n_{1}\cdot OPL_{2}}{c}-\frac{OPL}{c}
\]

Through this new time we get the geometric fluctuation of the delay time $\Delta t_{G}(r_{0},D)$:

\begin{equation}
	\Delta t_{G}(r_{0},D)=\left|t_{2,G}(r_{0},D)-t_{1}\right|
\end{equation}

Taking into account that:

	\[t_{1}=\frac{n_{1}\cdot OPL_{1}}{c}-\frac{OPL}{c}
\]

we obtain the following formula:

\begin{equation}
\label{delay8}
	\Delta t_{G}(r_{0},D)=\frac{n_{1}}{c}\cdot \left|OPL_{2}-OPL_{1}\right|.
\end{equation}

\subsection{Physical delay time fluctuation}

In this Section, we calculate the delay time ($t_{2,P}(r_{0},D)$) due to the physical variation of the atmosphere.
In the previous section we saw how the OPL changes due to the atmosphere (image motion). The change in OPL also induces a change in the refractive index. 
In fact, fluctuations in optical path  length must correspond to refractive angle variations.
We can calculate the refractive angle fluctuation assuming the atmosphere uniform distribution:

\begin{equation}
	\Delta\theta=arcos\left[\frac{OPL_{1}}{\sqrt{OPL^{2}_{1}+\Delta^{2}}}\right]
\end{equation}

The corresponding atmospheric refractive index $n_{2}$, obtained by Snell's law, is:

\begin{equation}
\label{index7}
	n_{2}(r_{0},D)=\frac{n_{1}sin\theta_{1}}{sin(\theta_{1}+\Delta\theta)}
\end{equation}

This refractive index is a function of $r_{0}$ and the telescope diameter($D$).

\subsection{Calculation of physical delay time fluctuation }
\label{pdtf}

With the value of the refractive index $n_{2}$ (Equation \ref{index7}) we can calculate a new physical delay time as a function of the $r_{0}$ 
and the telescope diameter ($D$), through the formula:

	\[t_{2,P}(r_{0},D)=\frac{n_{2}\cdot OPL}{c/n_{2}}-\frac{OPL}{c}
\]

where OPL is given by the Formula \ref{opl}.
Then, as in Section \ref{gf}, we obtain the fluctuation due to this variation $\Delta t_{P}(r_{0},D)$:

\begin{figure}
  \centering
  \includegraphics[width=8.5cm]{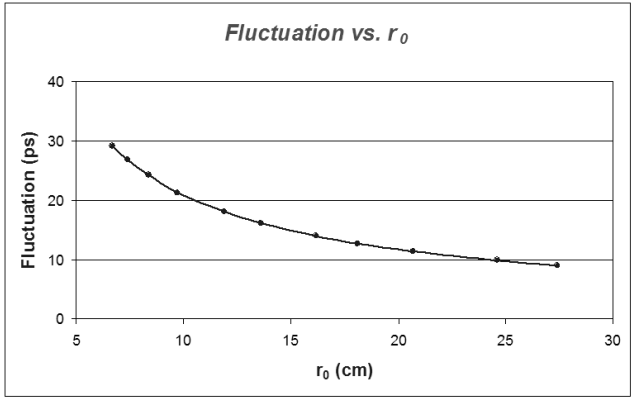}
  \caption{Fluctuation delay time as a function of the $r_{0}$. This simulation pertains to La Silla ($\lambda=0.632\mu m$, Zenith angle $=10^{\circ}$).}
             \label{r0}
   \end{figure}

\begin{equation}
	\Delta t_{P}(r_{0},D)=\left|t_{2,P}(r_{0},D)-t_{1}\right|
\end{equation}

where:

	\[t_{1}=\frac{n_{1}\cdot OPL_{1}}{c}-\frac{OPL}{c}
\]

we obtain:

\begin{equation}
\label{fis}
	\Delta t_{P}(r_{0},D)=\frac{OPL}{c}\cdot \left|n^{2}_{2}-n^{2}_{1}\right|.
\end{equation}

Through the Formula \ref{fis} we can estimate that a delay time fluctuation of 10.0 ps corresponds to a refractive index variation $\Delta n$ of $3.3\times 10^{-8}$. This gives an idea of the error propagation.

\begin{table}
 \centering
 \begin{minipage}{80mm}
 \caption{Fluctuation vs Zenith angle for different sites ($\lambda=0.632\mu m$, $r_{0}=15cm$).}
  \label{dia1}
  \begin{tabular}{@{}llllcccccccc@{}}
  \hline
 Zenith Angle & $\left(^{\circ}\right)$ & 0 & 15 & 30 & 45 & 60 \\

 \hline
 La Silla (3.58m)   & (ps)  & 9.8 & 10.4 & 12.4 & 16.9 & 27.5 \\
 Paranal   (8.20m) & (ps) & 8.2 & 8.7 & 10.4 & 14.2 & 23.1  \\
 Armazones (42.00m) & (ps)  & 5.8 & 6.1 & 7.3 & 9.9 & 16.2 \\
 \hline

\end{tabular}
\end{minipage}
\end{table}

\begin{figure}
  \centering
  \includegraphics[width=8.5cm]{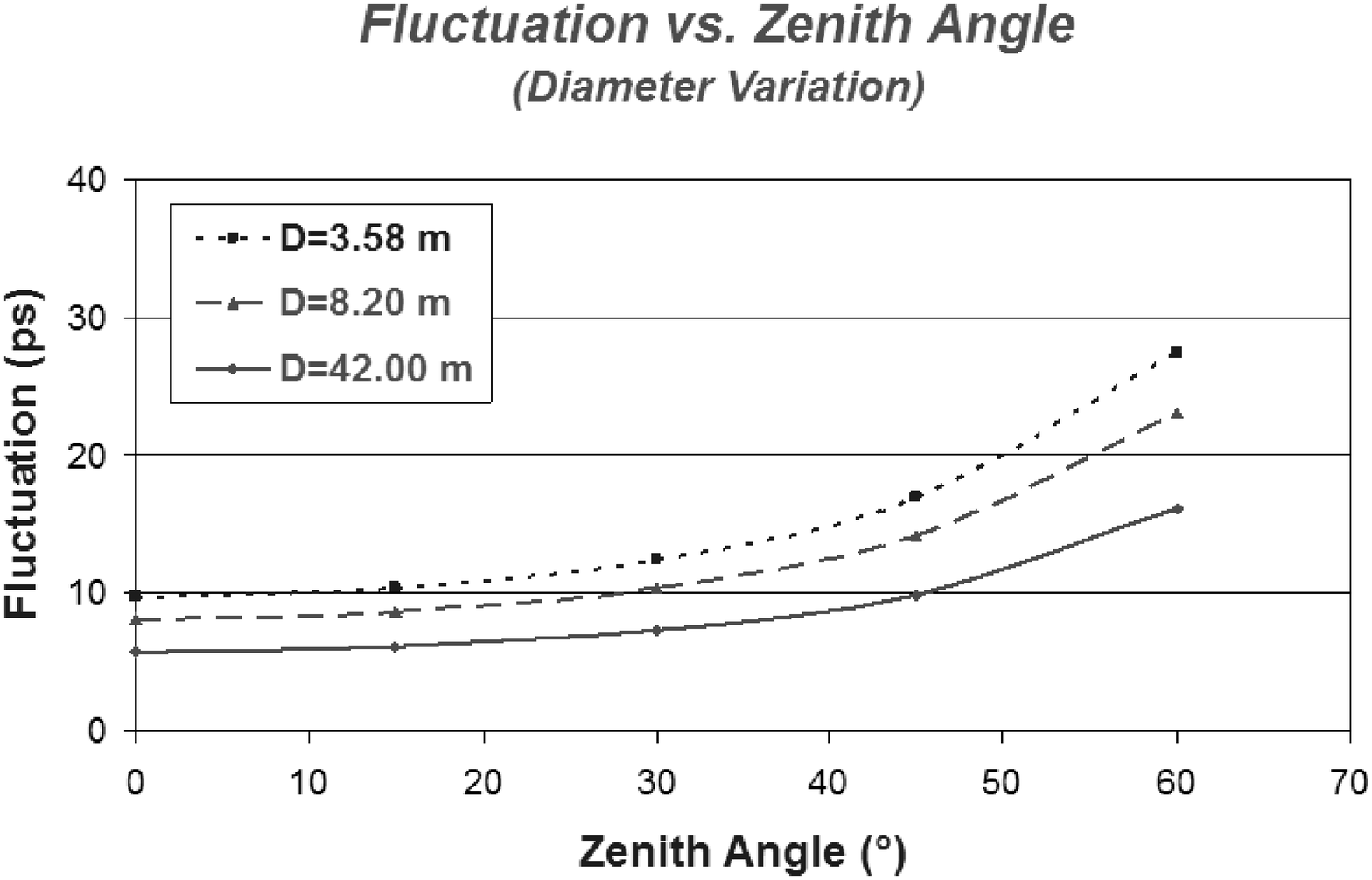}
  \caption{Delay time fluctuation as a function of the telescope diameter and the Zenith angle variation ($\lambda=0.632\mu m$, $r_{0}=15cm$). The simulation compares La Silla (NTT), Paranal (VLT) and Armazones (ELT).}
             \label{dia2}
   \end{figure}

\section{Calculation of delay time fluctuations}

Consequently, the total fluctuation of the delay time is the sum of the geometric and physical component:

	\[\Delta t_{F}=\Delta t_{G}+\Delta t_{P}
\]

In this sum the physical component is larger than the geometric component of four orders of magnitude.
Finally, we express this fluctuation through the extended formula:

\[\Delta t_{F}(r_{0},D)= \frac{n^{2}_{1}\cdot OPL}{c}\left|\sqrt{2}sin\sqrt{K\cdot \lambda^{\epsilon}\cdot D^{-\frac{\alpha}{\beta}}\cdot r^{-\frac{\gamma}{\eta}}_{0}}-1\right|+
\]

\begin{equation}
	+\frac{OPL}{c}\cdot\left|\left[\frac{n_{1}sin\theta_{1}}{sin(\theta_{1}+\Delta\theta)}\right]^{2}-n^{2}_{1}\right|
\end{equation}

where:

	\[\Delta\theta=arcos\left[\left(2sin^{2}\sqrt{K\cdot \lambda^{\epsilon}\cdot D^{-\frac{\alpha}{\beta}}\cdot r^{-\frac{\gamma}{\eta}}_{0}}+1\right)^{-\frac{1}{2}}\right]
\]

Substituting the constants with these values (Roddier, \cite{roddier}) $K=0.18$, $\alpha=1$, $\beta=3$,
$\gamma=5$, $\eta=3$ and $\epsilon=2$ we obtain the final formula used in the model:

	\[\Delta t_{F}(r_{0},D)= \frac{n^{2}_{1}\cdot OPL}{c}\left|\sqrt{2}sin\sqrt{0.18\cdot \lambda^{2}\cdot D^{-\frac{1}{3}}\cdot r^{-\frac{5}{3}}_{0}}-1\right|+
\]

\begin{equation}
	+\frac{OPL}{c}\cdot\left|\left[\frac{n_{1}sin\theta_{1}}{sin(\theta_{1}+\Delta\theta)}\right]^{2}-n^{2}_{1}\right|
	\label{cava}
\end{equation}

where:

	\[\Delta\theta=arcos\left[\left(2sin^{2}\sqrt{0.18\cdot \lambda^{2}\cdot D^{-\frac{1}{3}}\cdot r^{-\frac{5}{3}}_{0}}+1\right)^{-\frac{1}{2}}\right].
\]

We note that this delay time fluctuation depends on the wavelength and on the telescope diameter. In particular, it also depends on the Fried radius, Figure \ref{r0} shows the trend of this fluctuation as a $r_{0}$ function. The simulation is done for La Silla ($\lambda=0.632\mu m$, Zenith angle $=10^{\circ}$).

\begin{figure}
  \centering
  \includegraphics[width=8.5cm]{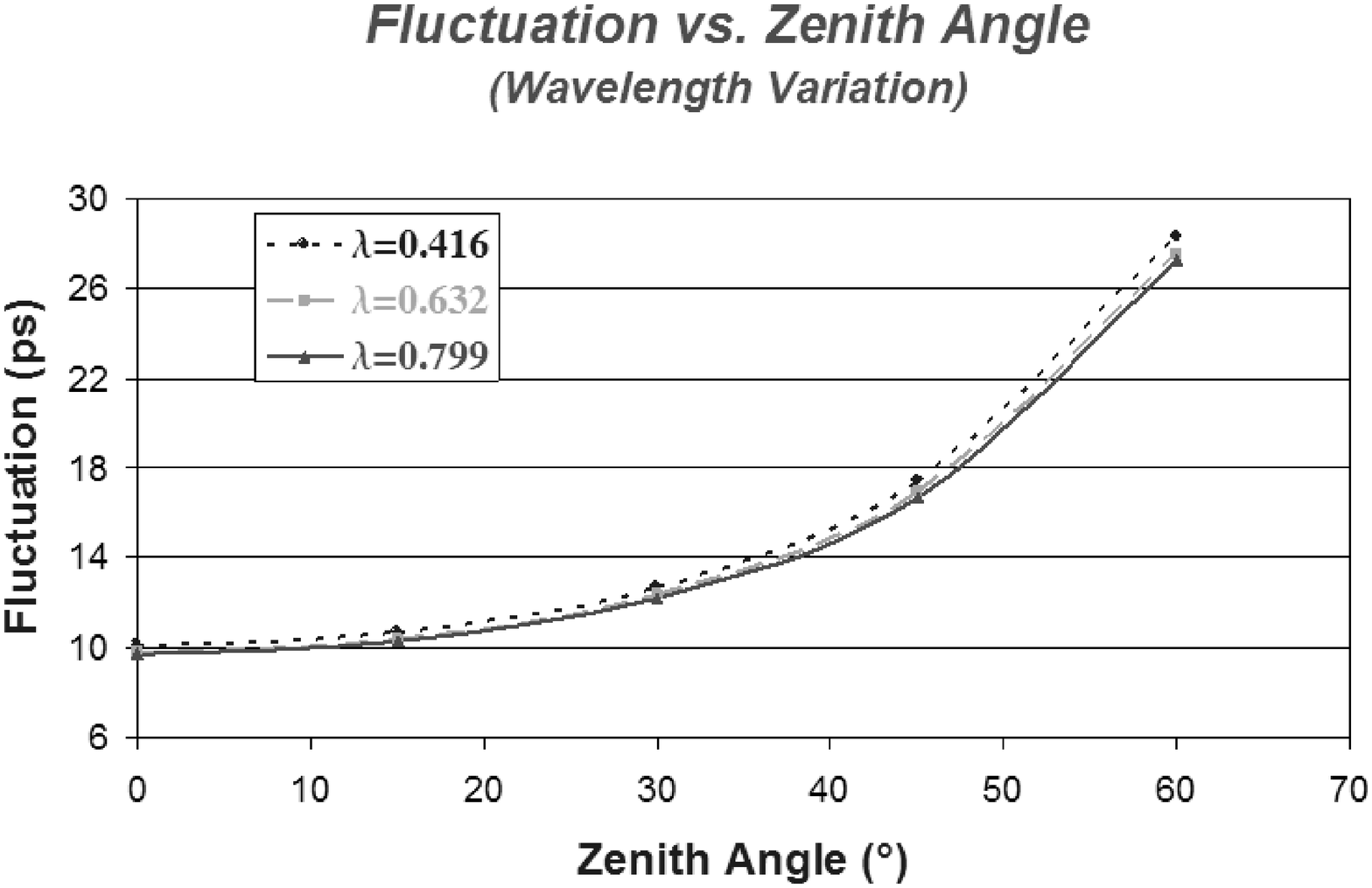}
  \caption{Delay time fluctuation as a function of the wavelength and the Zenith angle variation. The simulation pertains to La Silla ($D=3.58m$, $r_{0}=15cm$).}
             \label{wave}
   \end{figure}

\subsection{Application of the model to ESO astronomical sites for the delay time fluctuations}

\begin{figure}
  \centering
  \includegraphics[width=8.5cm]{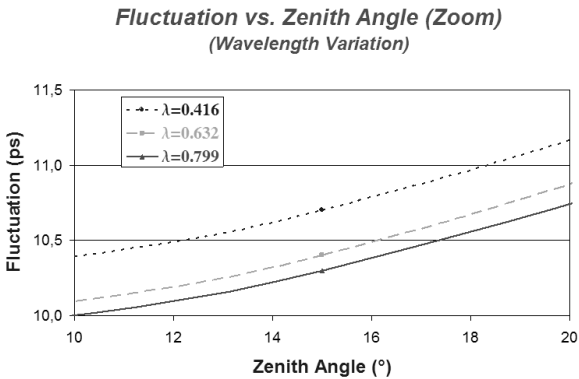}
  \caption{Zoom of Figure \ref{wave}.}
             \label{zoom}
   \end{figure}

In this Section we apply the previously described model to the three Chilean sites of ESO telescopes. We present the results obtained through the Formula \ref{cava}. Table \ref{dia1} shows the simulation results for each site, in particular the fluctuations variation as a function of the telescope diameter. We note that the fluctuation decreases with a larger diameter. In this case, between La Silla and Armazones telescopes the delay time fluctuations decreased of about 40\%. Figure \ref{dia2} shows the results of this model for each site as a function of Zenith angle.\\
Table \ref{wa} shows the simulation results for three wavelengths. In this case the simulation is done for La Silla. Figure \ref{wave} shows the Table \ref{wa} results and Figure \ref{zoom} is the zoom of these trends. In fact, we note that in the range $\lambda=0.416-0.799\mu m$ the fluctuation changes only by 4\%.

\begin{table}
 \centering
 \begin{minipage}{80mm}
 \caption{Fluctuation vs Zenith angle for different wavelengths, La Silla ($D=3.58m$, $r_{0}=15cm$).}
   \label{wa}
  \begin{tabular}{@{}llllcccccccc@{}}
  \hline
 Zenith Angle & $\left(^{\circ}\right)$ & 0 & 15 & 30 & 45 & 60 \\

 \hline
 $\lambda=0.416\mu m$  & (ps)  & 10.1 & 10.7 & 12.7 & 17.4 & 28.3 \\
 $\lambda=0.632\mu m$ & (ps) & 9.8 & 10.4 & 12.4 & 16.9 & 27.5  \\
 $\lambda=0.799\mu m$ & (ps)  & 9.7 & 10.3 & 12.2 & 16.7 & 27.3 \\
 \hline

\end{tabular}
\end{minipage}
\end{table}

\section{Inverting the model to calculate Fried radius}
\label{im2}

In this Section, we reverse the Formula \ref{cava} in order to derive a mathematical expression giving the $r_{0}$ value from the study of the delay time fluctuations (see Appendix \ref{im}). The $r_{0}$ value is expressed as a function of the telescope diameter (D), of the wavelength ($\lambda$) and of the variable ($\chi$).
This variable is obtained through the observation of the delay time fluctuations and the $n_{1}$ calculation. The model gives us the following formula for the Fried radius calculation:

\begin{equation}
	r_{0}=\left|0.54\cdot D^{-\frac{1}{5}}\cdot \lambda^{\frac{6}{5}}\cdot arctg^{-\frac{6}{5}}\left(\chi_{1,2}\right)\right|
\end{equation}

where:

	\[\chi_{1,2}=\frac{-tg\theta_{1}\pm\sqrt{tg^{4}\theta_{1}(Z-1)+ tg^{2}\theta_{1}\cdot Z}}{Z-tg^{2}\theta_{1}}
\]

and:

	\[Z=\frac{c\cdot\Delta t_{F}(r_{0},D)+n^{2}_{1}\cdot OPL}{n^{2}_{1}\cdot OPL}
\]

Through this model we have a new way to calculate the Fried radius as a function of the OPL and average refraction index of the atmosphere. Table \ref{simu1} shows a simulation for La Silla, the mathematical details are given in Appendix \ref{im}.

\begin{table}
 \centering
 \begin{minipage}{80mm}
  \caption{Simulation of the $r_{0}$ calculation through the delay time fluctuations for La Silla ($\lambda=0.632\mu m$, telescope diameter $D=3.58m$).}
   \label{simu1}
  \begin{tabular}{@{}lcc@{}}
  \hline
           Delay time fluctuation  & Fried radius \\
  \hline
         $\Delta t_{P}(r_{0},D)$ & $r_{0}$ \\
           ps &     cm    \\
 \hline
 
       $27.0$   &  $5.0$   \\
       $18.2$  &  $10.0$     \\
       $10.2$  &  $20.0$    \\

 \hline

\end{tabular}
\end{minipage}
\end{table}

\section{Conclusion}

We have shown in this paper that the delay and dispersion introduced by the terrestrial atmosphere in photon arrival times is significant to the level of tens of nanoseconds and tens of picoseconds respectively.\\
These values mean a severe degradation of the time tagging performances of modern astronomical detectors such as the Avalanche Photodiode (e.g. Barbieri \cite{barbieri} and Naletto \& Barbieri \cite{naletto}).
Furthermore, they imply that the very accurate timing signals available e.g. via radio signals by GPS or other Satellite Navigation System are not exploited to their full extent in the astrophysical field, as instead is done in geodetic satellite ranging applications.  The scientific results of very high time resolution astrophysics can therefore be degraded by such neglect. In particular, the E-ELT will provide a 25-fold increase of photon flux over existing telescopes thus opening the way to 'quantum' astronomy (e.g. Dravins et al. \cite{dravins}).  
The algorithms expounded in the present paper overcome these limitations.\\ 
In Section \ref{im2} we have described a theoretical mathematical model for calculating the Fried radius through the study of delay time fluctuations. Table \ref{simu1} shows the simulation results for La Silla. This is a completely new method for the study of atmospheric turbulence, in future work we want to correlate the delay time fluctuations experimental data with the traditional techniques for the Fried radius calculation. This will allow us a broader view of Earth's atmosphere and its influence on propagation of photons.

\subsection{ACKNOWLEDGMENTS}

Work partly supported by the University of Padova through 'Quantum Future', a strategic program started in 2008.  
Thanks are due to Prof. D. Dravins for useful discussions.

\appendix

\section{Inverting the model}
\label{im}

We start from Equation \ref{cava} relating the $\Delta t_{F}(r_{0},D)$ to the Fried parameter $r_{0}$:

\[\Delta t_{F}(r_{0},D)= \frac{n^{2}_{1}\cdot OPL}{c}\left|\sqrt{2}sin\sqrt{0.18\cdot \lambda^{2}\cdot D^{-\frac{1}{3}}\cdot r^{-\frac{5}{3}}_{0}}-1\right|+
\]

	\[+\frac{OPL}{c}\cdot\left|\left[\frac{n_{1}sin\theta_{1}}{sin(\theta_{1}+\Delta\theta)}\right]^{2}-n^{2}_{1}\right|
\]

where:

\begin{equation}
\label{dt1}
\Delta\theta=arcos\left[\left(2sin^{2}\sqrt{0.18\cdot \lambda^{2}\cdot D^{-\frac{1}{3}}\cdot r^{-\frac{5}{3}}_{0}}+1\right)^{-\frac{1}{2}}\right]
\end{equation}

In the case of studying the delay time fluctuations for the $r_{0}$ calculation, the geometric delay time fluctuation becomes negligible.\\
Then we consider the formula:

\begin{equation}
	\Delta t_{F}(r_{0},D)=\frac{OPL}{c}\cdot\left|\left[\frac{n_{1}sin\theta_{1}}{sin(\theta_{1}+\Delta\theta)}\right]^{2}-n^{2}_{1}\right|
\end{equation}

we isolate the trigonometric functions:

	\[\frac{c\cdot\Delta t_{F}(r_{0},D)+n^{2}_{1}OPL}{n^{2}_{1}OPL}=\frac{sin^{2}\theta_{1}}{sin^{2}(\theta_{1}+\Delta\theta)}
\]

we substitute:

\begin{equation}
Z=\frac{c\cdot\Delta t_{F}(r_{0},D)+n^{2}_{1}\cdot OPL}{n^{2}_{1}\cdot OPL}	
\end{equation}

Using the trigonometric addition formulas and parameters of the $sin$ and $cos$ functions we get:

	\[(Z-tg^{2}\theta_{1})\chi^{2}+2\cdot Z\cdot tg\theta_{1}\chi+Z\cdot tg^{2}\theta_{1}-tg^{2}\theta_{1}=0
\]

where:

\begin{equation}
\label{dt2}
	\chi=tg\Delta\theta
\end{equation}

and solving in the same variable we obtain:

\[\chi_{1,2}=\frac{-tg\theta_{1}\pm\sqrt{tg^{4}\theta_{1}(Z-1)+ tg^{2}\theta_{1}\cdot Z}}{Z-tg^{2}\theta_{1}}
\]

We now consider the equations \ref{dt1} and \ref{dt2} obtaining the following relationship:

\begin{equation}
\label{dt3}
	\Delta\theta=arctg \chi=arcos\left[\left(2sin^{2}\sqrt{0.18\lambda^{2}D^{-\frac{1}{3}}r^{-\frac{5}{3}}_{0}}+1\right)^{-\frac{1}{2}}\right]
\end{equation}

Finally, from equation \ref{dt3} we explicitly the $r_{0}$ value and we obtain the following formula:

\begin{equation}
	r_{0}=\left|0.54\cdot D^{-\frac{1}{5}}\cdot \lambda^{\frac{6}{5}}\cdot arctg^{-\frac{6}{5}}\left(\chi_{1,2}\right)\right|.
\end{equation}

\label{lastpage}

\end{document}